\documentstyle[aps]{revtex}
\begin{document}
\def\hrf{\hrulefill}
\def\A{{\rm A}}
\def\c{\centerline}
\def\v{\vskip 1pc}
\def\ej{\vfill\eject}
\def\r{\vec r}
\def\ra{\rangle}
\def\la{\langle}
\def\re{{\rm e}}
\def\tr{{\rm tr}}
\def\half{{1 \over 2}}
\def\da{\dagger}
\def\pup{{\rm p} \uparrow}
\overfullrule 0 pc
\title{{\bf Temporal discretisation of }\\
{\bf the Skyrme Model}}
\author{George Jaroszkiewicz$^1$ and Vladimir Nikolaev$^2$}
\address{$^1$School of Mathematical Sciences, University of Nottingham, University\\
Park, Nottingham NG7 2RD, UK;\\
$^2$Bulgarian Academy of Sciences,\\
Institute of Nuclear Research and Nuclear Energy,\\
Theoretical Nuclear Physics Group, 72\\
Tzarigradsko Shousse Blvd \\
Sofia 1784, Bulgaria}
\maketitle
\section{Introduction}
Throughout this paper {\it CT} and {\em DT }refer to {\it
continuous time} and {\it discrete time }respectively, units are
taken as $c=\hbar =1$ and the metric tensor has components $\eta
_{\mu \nu }=$ diag$\left( 1,-1,-1,-1\right) $.$\;$The symbol $=_c$
denotes an equality holding modulo the equations of motion.
In recent years there has been much interest in the discretization of time
for a variety of reasons. These include the study of
{\bf integrable systems} \cite{MOSER+VESELOV-91}, motivated by the question
of the integrability of various dynamical models. Systems studied are often
point particle models with special properties which make the process of
discretization successful, such as the Toda lattice \cite{SURIS-97B};
{\bf field theories} [3-4], considered as approximations to CT field
theories. Generally these tend to look like variants of lattice gauge
theory, with the important difference that time is not Wick rotated;
{\bf lattice gauge theory}, which has as its objective the CT limit $%
a\rightarrow 0$, where $a$ is the lattice spacing and which is regarded as
the place where physics occurs. The value $a\neq 0$ is taken only as a
calculational device, suitable for providing a regularization procedure and
for computer simulation to some given order of numerical accuracy. In this
scheme discretization is also applied to the spatial coordinates, and
because time is Euclidean, the theory uses a four dimensional Euclidean
space.
On a more fundamental level it has been speculated by many authors that
perhaps space and/or time are not continuous on very short scales. In the
history of quantum field theory, problems associated with operator products
at the same spacetime point were sometimes addressed by ``point-splitting''
techniques, wherein two or more field operators at the same point were
defined at separate points and only after the calculations was the
separation taken to zero. This was a technique used for example by Schwinger
in his famous 2-dimensional model of QED \cite{SCHWINGER-62}. More recently
there has been a strong suspicion that at the Planck scale the usual view of
spacetime breaks down and novel new ideas should be contemplated.
In a series of papers [3-6] \nocite{JAROSZKIEWICZ-97A} \nocite
{JAROSZKIEWICZ-97B} \nocite{JAROSZKIEWICZ-97C} \nocite{JAROSZKIEWICZ-97D}
attention was focussed on the consequences of discretization of time but not
of space. Such an idea was discussed by 't Hooft in $2+1$ dimensions \cite
{T'HOOFT-96}, with the suggestion that under certain circumstances, there
may be variants of the Regge calculus approach to General relativity \cite
{REGGE-61} wherein time becomes discrete but space does not. More recently,
discrete time has been discussed as a direct consequence of quantum
processes in action \cite{JAROSZKIEWICZ-01}. In this approach, time jumps
whenever a factor state in the quantum state of the Universe undergoes a
test resulting in information gain. These factor states can in principle
include attributes such as momentum, so there is no necessity to discuss
discrete space as well, although that is possible. In this article only
temporal discretization will be discussed.
In the system of DT mechanics discussed here, it is considered an exact form
of mechanics, with its own consistent laws of motion and dynamical
invariants, and not only in the CT limit. This is a fundamental difference
between this DT mechanics and conventional lattice gauge theory. All of the
operators in our DT quantum theory have to be good, in the language of
lattice gauge theory.
It is the case that a{\it ny} conventional continuous time theory can be
discretized, simply by looking at the system at discrete times. For example,
integrating the Lagrangian between chosen times (temporal nodes) gives
Hamilton's principal function (the Hamilton-Jacobi function), which can be
regarded as a convenient discretization of time, because it depends only on
what the system is doing (the co-ordinates) at the end points of the
temporal interval.
Our original motivation for discretizing time arose when computer simulation
of a soliton theory was undertaken \cite{JAROSZKIEWICZ-95B}. In CT theory,
various integrals such as total mass, linear and angular momentum and charge
remain invariants under dynamical evolution. However, it was soon found that
a naive discretization of the equations of motion did not lead to the naive
discretizations of these integrals remaining invariant during the
simulations. This has nothing to do with numerical inaccuracy, but with the
principles being employed. It was perceived that discretization should
perhaps be done in a more carefully controlled, principled way, so that a
genuine discrete dynamics with true invariants emerged.
In this paper some of the methodology discussed in \cite{JAROSZKIEWICZ-97B}
is applied to a well-known field theoretic model with an underlying
non-abelian group structure, the Skyrme model. This is not a trivial
application, for it is not obvious at all that every model in CT field
theory carries over easily into a DT analogue theory with DT analogues of
the CT structures remaining intact. An important difference is that there
may be no obvious DT analogue of the CT Hamiltonian. It is the merit of
integrable systems that such an analogue should be found, but in general
this will not be so. The obvious reason is that by definition, there is no
continuous translation in time in DT mechanics, so there is no DT analogue
of the CT generator of such a transformation.
Fortunately, the basic idea of a Noether invariant survives in DT mechanics.
If there is a continuous symmetry of the system function (the DT analogue of
the CT Lagrangian) then there will be a corresponding invariant, referred to
here as a Maeda-Noether invariant \cite{MAEDA-81}. In the DT Skyrme model,
the symmetry which generates the axial and vector charges survives and the
corresponding DT charges are discussed here.
We shall also discuss T.D. Lee's version of {\em DT} mechanics, which
automatically gives a {\em DT} analogue of energy.
\section{DT mechanics}
In the models discussed here the dynamical variables are defined only at
instants of time $nT$ where $n$ is an integer and $T$ is some fixed time
interval. In these models there are no gauge fields defined on the temporal
links connecting successive instants of time, so the formalism presented
here takes on a simpler form than would be the case in say QED or QCD.
The dynamical variables here are real valued fields $\varphi ^\alpha
,\;\alpha =0,1,2,3$ with suitable constraints. The index $\alpha $ is not a
Lorentz index, but it is lowered with the metric tensor components $\eta
_{\mu \nu }$. The DT analogue of the CT Lagrange density ${\cal L=L}\left(
\varphi ^\alpha ,\partial _\mu \varphi ^\alpha \right) $ is called the {\em %
system function density} and takes the form
\begin{equation}
{\cal S}^n\equiv {\cal S}\left( \varphi _n^\alpha ,\varphi _{n+1}^\alpha
,\nabla \varphi _n^\alpha ,\nabla \varphi _{n+1}^\alpha \right) .
\end{equation}
It has the physical dimensions of an action density, not an energy density.
The integral $S^n\equiv \int d{\bf x}{\cal S}^n$ is called the {\em system
function}, and although used as the {\em DT} analogue of the Lagrangian $L$
in {\em CT} mechanics, is more like a Hamilton's principal function.
Applying Hamilton's principle to the action sum
\begin{equation}
A_{MN}\left[ \Gamma \right] =\sum_{n=M}^{N-1}S^n
\end{equation}
gives the second order equations of motion \cite{JAROSZKIEWICZ-97B}
\begin{equation}
\Pi _n^\alpha =_{c} \bar{\Pi}_n^\alpha ,\;\;\;\;M<n<N-1
\label{eq}
\end{equation}
where
\begin{equation}
\Pi _n^\alpha \equiv -\frac{\partial {\cal S}^n}{\partial \varphi _n^\alpha }%
+\nabla \cdot \frac{\partial {\cal S}^n}{\partial \nabla \varphi _n^\alpha }%
,\;\;\;\bar{\Pi}_n^\alpha \equiv \frac{\partial {\cal S}^{n-1}}{\partial
\varphi _n^\alpha }-\nabla \cdot \frac{\partial {\cal S}^{n-1}}{\partial
\nabla \varphi _n^\alpha }.
\end{equation}
Given an infinitesimal transformation of the fields $\varphi ^\alpha
\rightarrow \varphi ^\alpha +\delta \varphi ^\alpha $ then
\begin{equation}
\delta {\cal S}^n{=}_{c}\delta \varphi _{n+1}^\alpha \bar{\Pi}%
_{n+1}^\alpha -\delta \varphi _n^\alpha \Pi _n^\alpha +\nabla
\cdot {\mathbf \sigma ,}
\end{equation}
where
\begin{equation}
\delta {\mathbf \sigma }_n\equiv \delta \varphi _n^\alpha \frac{\partial
{\cal S}%
^n}{\partial \nabla \varphi _n^\alpha }+\delta \varphi _{n+1}^\alpha \frac{%
\partial {\cal S}^n}{\partial \nabla \varphi _{n+1}^\alpha }.
\end{equation}
A symmetry of the system function is a transformation of the fields which
leaves the system function unchanged. Hence for such a symmetry
\begin{eqnarray}
&&\delta \varphi _{n+1}^\alpha \bar{\Pi}_{n+1}^\alpha -\delta \varphi
_n^\alpha \Pi _n^\alpha +\nabla \cdot \delta {\mathbf \sigma }_n=0,\;\;
\nonumber \\
\Rightarrow \;\;\; &&\delta \varphi _{n+1}^\alpha \Pi _{n+1}^\alpha -\delta
\varphi _n^\alpha \Pi _n^\alpha +\nabla \cdot \delta {\mathbf \sigma }%
_n=_c0,\;
\end{eqnarray}
which can be written in the forwards form
\begin{equation}
D_n^{+}\delta \rho _n+\nabla \cdot \delta {\bf j}_n=_c0
\label{cons}
\end{equation}
using the equations of motion $\left( \ref{eq}\right) .\;$Here the forwards
densities are defined by
\begin{equation}
\delta \rho _n\equiv \delta \varphi _n^\alpha \Pi _n^\alpha ,\;\;\;\delta
{\bf j}_n\equiv T^{-1}\delta {\mathbf \sigma }_n,
\end{equation}
with $D_n^{+}\equiv \left( U_n-1\right) /T$ , where $U_n$ is the temporal
step operator defined by $U_nf_n=f_{n+1}$. From $\left( \ref{cons}\right) $
the corresponding dynamical invariant is given by
\begin{equation}
\delta C^n\equiv \int \delta \varphi _n^\alpha \Pi _n^\alpha ,
\end{equation}
assuming that the fields fall off sufficiently rapidly at spatial infinity.
An alternative form is given by noting that
\begin{equation}
\;\delta \varphi _{n+1}^\alpha \bar{\Pi}_{n+1}^\alpha -\delta \varphi
_n^\alpha \bar{\Pi}_n^\alpha +\nabla \cdot \delta {\mathbf \sigma }%
_n=_c0,\;\;\;\;
\end{equation}
giving the backwards DT equation of continuity
\begin{equation}
D_n^{-}\delta \bar{\rho}_n+\nabla \cdot \delta {\bf j}_n=_c0
\end{equation}
where $D_n^{-}\equiv \left( 1-U_n^{-1}\right) /T$ and the backwards
densities are defined by
\begin{equation}
\delta \bar{\rho}_n\equiv \delta \varphi _{n+1}^\alpha \bar{\Pi}%
_{n+1}^\alpha ,\;\;\;\;\;\delta {\bf j}_n\equiv T^{-1}\delta {\mathbf
\sigma }_n.
\end{equation}
\section{The Skyrme model}
The Skyrme model \cite{SKYRME-61}  has soliton
solutions with a number of properties suggestive of baryon physics. Its
basic dynamical degrees of freedom are space-time fields $U\left( x\right) $
which take values in $SU(2).$ Because this implies constraints, it is often
convenient to parametrize these fields in terms of an unconstrained isotopic
triplet of real scalar fields ${\mathbf \pi }\equiv (\pi ^1,\pi ^2,\pi
^3):$%
\begin{equation}
U\equiv \exp \left\{ i{\mathbf \tau }{ \cdot }{\mathbf \pi
}\right\} =\cos \left( \mid {\mathbf \pi }|\right)
+i\frac{{\mathbf \tau }\cdot {\mathbf
\pi }}{|{\mathbf %
\pi }|}\sin \left( \mid {\mathbf \pi }|\right) ,  \label{q23}
\end{equation}
where ${\mathbf \tau }\equiv \left( \tau ^1,\tau ^2,\tau ^3\right) $ are
the
Pauli matrices. With the definitions
\begin{equation}
U_\mu \equiv \partial _\mu U,\;\;\;\;L_\mu \equiv U^{+}\partial _\mu
U=-\partial _\mu U^{+}U
\end{equation}
the Lagrange density may be written in the form
\begin{equation}
{\cal L}_s\equiv -\frac{F_\pi ^2}{16}Tr\,L_\mu L^\mu -\frac
1{32e^2}Tr\,\left[ L_\mu ,L_\nu \right] \left[ L^\mu ,L^\nu \right]
\end{equation}
where $F_\pi $ is the pion coupling constant and the second term is known as
the {\it Skyrme term}. In terms of the $U$ fields this is equivalent to
\begin{equation}
{\cal L}_s=\frac{F_\pi ^2}{16}Tr\,U_\mu ^{+}U^\mu -\frac 1{16e^2}Tr\left\{
U_\mu ^{+}U_\nu U^{+\mu }U^\nu -U_\mu ^{+}U_\nu U^{+\nu }U^\mu \right\} .
\end{equation}
In addition to the standard Poincar\'{e} symmetries an important symmetry of
this Lagrangian is invariance under separate left and right $SU(2)$
transformations;
\begin{equation}
U\rightarrow U^{\prime }\equiv AUB^{+}
\end{equation}
where $A$ and $B$ are spacetime independent elements of $SU(2).$ This
generates the so-called axial and vector charges.
In Appendix A the quaternion approach to the parametrization of the $U$
variables is given. With this, $U$ may be written in the form
\begin{equation}
U=q_\mu \varphi ^\mu =\varphi ^0+i\tau ^i\varphi ^i,
\end{equation}
where $q_0\equiv I_2,$ the $2\times 2$ identity matrix and $q_i\equiv i\tau
^i,$ $i=1,2,3$ have the properties of the quaternions $i,\;j,\;k$. The four
real fields $\varphi ^\mu $ are read off from $\left( \ref{q23}\right) $ to
be
\begin{equation}
\varphi ^0\equiv \cos \left( \mid {\mathbf \pi
}|\right) ,\;\;\;\;\;\varphi
^i\equiv \sin \left( \mid {\mathbf \pi }|\right) n^i{\bf
,\;\;\;\;\;\;n\cdot n=1.%
}
\end{equation}
Since there are only three independent parameters describing the elements of
$SU\left( 2\right) $, the four components $\varphi ^\mu $ are constrained to
the surface of $S^3,$ the unit sphere in four dimensions, i.e.
\begin{equation}
\varphi ^\mu \varphi ^\mu =1.  \label{1}
\end{equation}
With this reparametrisation the Lagrange density becomes
\begin{equation}
{\cal L}_s=\frac{\alpha ^2}2\partial _\mu \varphi ^\alpha \partial ^\mu
\varphi ^\alpha -\frac{\beta ^2}4\partial _\mu \varphi ^\alpha \partial
_v\varphi ^\beta \left\{ \partial ^\mu \varphi ^\beta \partial ^v\varphi
^\alpha -\partial ^\mu \varphi ^\alpha \partial ^\nu \varphi ^\beta \right\}
+\frac{{}_1}{{}^2}\mu \left( \varphi ^\alpha \varphi ^\alpha -1\right)
\end{equation}
where $\alpha ^2\equiv \frac{_1}{^4}F_\pi ^2,\;\beta ^2\equiv e^{-2}$ and
the Lagrange multiplier $\mu $ enforces the $S^3$ constraint $\left( \ref{1}%
\right) .$ Then the conjugate momenta are given by
\begin{equation}
\pi ^\alpha \equiv \frac{\partial {\cal L}}{\partial \dot{\varphi}^\alpha }%
=M_{\alpha \beta }\dot{\varphi}^\beta
\end{equation}
where
\begin{equation}
M_{\alpha \beta }=\left( \alpha ^2-\beta ^2\partial _i\varphi ^\mu \partial
_i\varphi ^\mu \right) \delta _{\alpha \beta }+\beta ^2\partial _i\varphi
^\alpha \partial _i\varphi ^\beta .
\end{equation}
The constraints turn out to be second class in the terminology of Dirac \cite
{DIRAC:64} and given by
\begin{equation}
\chi _1\equiv \varphi ^\alpha \varphi ^\alpha -1\approx 0,\;\;\;\chi
_2\equiv \varphi ^\alpha \pi ^\alpha \approx 0
\end{equation}
Then the non-zero Dirac brackets are evaluated to be
\begin{eqnarray}
\left\{ \pi _{{\bf x}}^\alpha ,\varphi _{{\bf y}}^\beta \right\} _{DB}
&=&\left( \varphi _{{\bf x}}^\alpha \varphi _{{\bf x}}^\beta -\delta
_{\alpha \beta }\right) \delta ^3\left( {\bf x-y}\right) ,  \nonumber \\
\left\{ \pi _{{\bf x}}^\alpha ,\pi _{{\bf y}}^\beta \right\} _{DB} &=&\left(
\pi _{{\bf x}}^\alpha \varphi _{{\bf x}}^\beta -\pi _{{\bf x}}^\beta \varphi
_{{\bf x}}^\alpha \right) \delta ^3\left( {\bf x-y}\right) .
\end{eqnarray}
It is these which should be used in the quantization of the fields.
\section{The SU(2) particle}
\subsection{continuous time}
In this section the most basic variant of the Skyrme model is considered,
which is to drop the Skyrme term and the spatial degrees of freedom. Then
the Lagrange density reduces to the Lagrangian
\begin{equation}
L=\frac{_1}{^4}\alpha ^2Tr\,\dot{U}^{+}\dot{U},  \label{22}
\end{equation}
where $U\equiv U\left( t\right) $ is a time dependent element of SU(2) and $%
\alpha =\frac{{}_1}{{}^2}F_\pi $. The number of independent real dynamical
variables is three and there are two alternative formulations:
\subsubsection{\protect\smallskip the $\pi $ fields}
The $U$ fields may be parametrized using three unconstrained real fields: $%
U\left( t\right) =\exp \left\{ i{\mathbf \tau }{ \cdot }{\mathbf
\pi }\left( t\right) \right\} ,$ where ${\mathbf \pi }\left(
t\right) \;\equiv F\left(
t\right) {\bf n}\left( t\right) $ is an element of ${\cal R}^3$ and ${\bf n}%
\left( t\right) $ is a unit $3-$vector. Then
\begin{equation}
U\left( t\right) =\cos F+i\sin F\,{\bf \tau }{ \cdot }{\bf n}.
\end{equation}
The mapping from the ${\cal R}^3$ space of the parameters ${\mathbf \pi
}\,$to $%
SU\left( 2\right) $ is many to one, with the vectors $\left( F+2k\pi \right)
{\bf n}$, $k$ an integer,\ mapping into the same point of $SU(2).$\ The
Lagrangian $\left( \ref{22}\right) $ then reduces to
\begin{equation}
L=\frac{{}_1}{{}^2}\alpha ^2\left\{ \dot{F}^2+{\bf \dot{n}}{ \cdot
}{\bf
\dot{n}}\sin ^2F\right\} +\frac{{}_1}{{}^2}\mu \left( {\bf n}{ \cdot }%
{\bf n}-1\right) ,
\end{equation}
where a Lagrange multiplier is included to enforce the normalization
condition on the unit vector ${\bf n}$. The equations of motion are
\begin{eqnarray}
\ddot{F}-\sin F\cos F\,{\bf \dot{n}\cdot \dot{n}} &&=_c0, \\
\alpha ^2\sin ^2F{\bf \ddot{n}}+2\alpha ^2\sin F\cos
F\,\dot{F}{\bf \dot{n}} &&=_c\mu {\bf n},\;\;\;{\bf n}{ \cdot
}{\bf n}{=}_{c} 1.
\end{eqnarray}
In phase space the system has two second class constraints in the language
of Dirac \cite{DIRAC:64}. Now define $p,\,{\bf p}$ to be the momenta
conjugate to $F$ and ${\bf n}$ respectively. Then these constraints take the
form
\begin{equation}
\chi _1\equiv \,{\bf n\cdot n}-1\approx 1,\;\;\;\chi _2\equiv {\bf n\cdot p}%
\;\approx 0.
\end{equation}
Following Dirac \cite{DIRAC:64}, the Dirac brackets can be constructed in
the standard way giving the non-zero brackets
\begin{equation}
\left\{ p,F\right\} _D=-1,\;\;\;\left\{ p^i,n^j\right\} _D=-\delta
_{ij}+n^in^j,\;\;\;\left\{ p^i,p^j\right\} _D=\,p^in^j-p^jn^i.
\end{equation}
The total Hamiltonian is given by
\begin{equation}
H_T=\frac{p^2}{2\alpha ^2}+\frac{{\bf p\cdot p}}{2\alpha ^2\sin ^2F},
\end{equation}
which is an invariant of the motion.
Two additional invariants of the motion can be found using Noether's theorem
by observing that the transformation
\begin{equation}
U\rightarrow U^{\prime }\equiv AUB^{+}
\end{equation}
is a symmetry of the Lagrangian, where $A$ and $B$ are space and time
independent elements of $SU\left( 2\right) $. Writing $A\simeq 1+i{\bf \tau }%
\cdot {\bf a,\;\;\;}B\simeq 1+i{\bf \tau }\cdot {\bf b,}$ where ${\bf a}$
and ${\bf b}${\bf \ }are infinitesimal then
\begin{equation}
\delta F=\left( {\bf a-b}\right) \cdot {\bf n,\;\;\;}\delta {\bf n}=\cot
F\,\left\{ {\bf a-b-n\cdot }\left( {\bf a-b}\right) {\bf n}\right\} +{\bf %
n\times }\left( {\bf a+b}\right)
\end{equation}
to lowest order in the infinitesimal parameters. Now an application of
Noether's theorem gives the conserved left and right charges
\begin{equation}
{\bf L}\,\equiv {\bf n\times p-}p{\bf n}-\cot F{\bf p,\;\;\;R}\equiv {\bf %
n\times p+}p{\bf n}+\cot F{\bf p}
\end{equation}
in phase-space. In configuration space they take the form
\begin{eqnarray}
{\bf L} &=&\alpha ^2\left\{ \sin ^2F{\bf n\times \dot{n}-}\dot{F}{\bf n}%
-\cos F\sin F{\bf \dot{n}}\right\} \equiv {\bf A}-{\bf V}  \nonumber \\
{\bf R} &=&\alpha ^2\left\{ \sin ^2F{\bf n}{ \times }{\bf \dot{n}+}\dot{F%
}{\bf n}+\cos F\sin F{\bf \dot{n}}\right\} \equiv {\bf A+V}
\end{eqnarray}
where
\[
{\bf A}\equiv \alpha ^2\sin ^2F{\bf n}{ \times }{\bf \dot{n},\;\;\;V}%
\equiv \alpha ^2\left( \dot{F}{\bf n}+\cos F\sin F{\bf \dot{n}}\right) ,
\]
are conserved separately. These are known conventionally as the{\bf \ }{\it %
vector} and {\it axial} charges respectively.
\subsubsection{The $\varphi $ fields}
The Lagrangian takes the form
\begin{equation}
L\equiv \frac{_1}{^2}\alpha ^2\dot{\varphi}^\alpha \dot{\varphi}^\alpha +%
\frac{_1}{^2}\mu \left( \varphi ^\alpha \varphi ^\alpha -1\right)
\end{equation}
and gives equations of motion
\begin{equation}
\alpha ^2\ddot{\varphi}^\alpha =_c\mu \varphi ^\alpha ,\;\;\;\varphi ^\alpha
\varphi ^\alpha =_c1,
\end{equation}
i.e.
\begin{equation}
\ddot{\varphi}^\alpha =_c\left( \varphi ^\beta \ddot{\varphi}^\beta \right)
\varphi ^\alpha .
\end{equation}
The Lagrangian is invariant to the global $SU(2)$ transformation $U^{\prime
}=AUB^{+}$ where $A\,$ and $B$ are time independent elements of $SU\left(
2\right) .$ Now suppose
\begin{equation}
A\simeq 1+i{\mathbf \tau }\cdot {\mathbf a},\;\;\;B\simeq 1+i{\bf \tau
}\cdot {\bf b}
\end{equation}
where ${\bf a}$ and ${\bf b}$ are infinitesimal, then
\begin{equation}
\delta \varphi ^0=\left( {\bf b-a}\right) \cdot {\mathbf \varphi
,\;\;\;}\delta
{\mathbf \varphi }=\left( {\bf a-b}\right) \varphi ^0+{\mathbf \varphi
}\times
\left( {\bf a+b}\right)   \nonumber
\end{equation}
Hence the conserved charge is given by
\[
\delta \rho =\left( {\bf a-b}\right) \cdot {\bf V+}\left( {\bf a+b}\right)
\cdot {\bf A}
\]
where
\begin{equation}
{\bf V}\equiv \alpha ^2\left( \varphi ^0{\mathbf
\dot{\varphi}-\dot{\varphi}}^0%
{\mathbf \varphi }\right) ,\;\;\;{\bf A}\equiv \alpha ^2{\mathbf
\dot{\varphi}}%
\times {\mathbf \varphi }
\end{equation}
with ${\bf \dot{V}}=_c$ ${\bf \dot{A}}=_c{\bf 0.}$
\subsection{temporal discretization}
Turning to the temporal discretization of the $SU(2)$ particle system, the
problem reduces to choosing a suitable virtual path between temporal notes
\cite{JAROSZKIEWICZ-97A}. It was found that appropriate paths were of the
form
\begin{equation}
U_{n\lambda }\equiv \lambda U_{n+1}+(1-\lambda )U_n
\end{equation}
where the parameter $\lambda \in \left[ 0,1\right] $ interpolates temporal
nodes and
\begin{equation}
U_n\equiv q_\alpha \varphi _n^\alpha ,\;\;\;\;\;\;\varphi _n^\alpha \varphi
_n^\alpha =1.
\end{equation}
Given the Lagrangian
\begin{equation}
L\equiv \frac{_1}{^4}\alpha ^2Tr\,\dot{U}^{+}\dot{U}
\end{equation}
then the system function becomes
\begin{equation}
S^n=\frac{_1}{^2}T\alpha ^2D_n^{+}\varphi _n^\alpha D_n^{+}\varphi _n^\alpha
+\frac{_1}{^4}\mu _nT\left( \varphi _n^\alpha \varphi _n^\alpha -1\right) +%
\frac{_1}{^4}\mu _{n+1}T\left( \varphi _{n+1}^\alpha \varphi _{n+1}^\alpha
-1\right)  \label{sys}
\end{equation}
The equations of motion are
\begin{equation}
\;\varphi _{n+1}^\alpha -2\varphi _n^\alpha +\varphi _{n-1}^\alpha =_c\mu
_nT^2\varphi _n^\alpha ,\;\;\;\varphi _n^\alpha \varphi _n^\alpha =_c1
\end{equation}
This is equivalent to
\begin{equation}
\varphi _{n+1}^\alpha +\varphi _{n-1}^\alpha =_c\varphi _n^\beta \left(
\varphi _{n+1}^\beta +\varphi _{n-1}^\beta \right) \varphi _n^\alpha
\end{equation}
There is invariance under
\begin{equation}
U_n\rightarrow U_n^{\prime }\equiv AU_nB^{+},\;\;\;\;\;A,B\in SU(2)
\end{equation}
Then if
\begin{equation}
A\simeq 1+i{\mathbf \tau }\cdot {\bf a} \;\;\;\; B\simeq 1+i{\mathbf \tau
\cdot b%
}
\end{equation}
then
\begin{equation}
\delta \varphi _n^0=-\left( {\bf a-b}\right) \cdot \bbox{\varphi }%
_n,\;\;\;\delta \bbox{\varphi}_n=\varphi _n^0\left( {\bf
a-b}\right) +\bbox{ %
\varphi }_n\times \left( {\bf a+b}\right)  \label{456}
\end{equation}
Then
\begin{equation}
\delta \rho _n\equiv \delta \varphi _n^\alpha \Pi _n^\alpha =\left( {\bf a-b}%
\right) \cdot {\bf V}^n+\left( {\bf a+b}\right) \cdot {\bf A}^n,
\end{equation}
giving the conserved charges
\begin{eqnarray}
{\bf A}^n &\equiv &\alpha ^2D_n^{+}\bbox{ \varphi }_n\times \bbox{
\varphi }_n=%
\frac{\alpha ^2}T\bbox{ \varphi }_{n+1}\times \bbox{ \varphi }_n
\nonumber \\
{\bf V}^n &\equiv &\alpha ^2\left[ \varphi _n^0D_n^{+}\bbox{ \varphi }%
_n-D_n^{+}\varphi _n^0\,\bbox{ \varphi }_n\right] =\frac{\alpha
^2}T\left( \varphi _n^0\bbox{ \varphi }_{n+1}-\varphi
_{n+1}^0\bbox{ \varphi }_n\right)
\end{eqnarray}
If these charges are non-zero, then there is a natural time independent
frame in isospace given by the directions (${\bf V}^n,{\bf A}^n,{\bf A}%
^n\times {\bf V}^n).$
The following argument simplifies the equations of motion and establishes
the existence of an infinite hierarchy of quadratic invariants. First note
that, regardless of the equations of motion, the quantity $U_{n+1}^{+}U_n$
is an element of $SU\left( 2\right) ,$ and so may be written in the form
\begin{equation}
U_{n+1}^{+}U_n=q_\lambda \Phi _n^\lambda
\end{equation}
using the quaternionic notation discussed in Appendix A, where
\begin{equation}
\Phi _n^\lambda \equiv c_{\,\,\,\,\,\,\,\,\,\,\,\lambda }^{\alpha \beta
}\varphi _{n+1}^\alpha \varphi _n^\beta
\end{equation}
and
\begin{equation}
\Phi _n^\lambda \Phi _n^\lambda =1.
\end{equation}
In detail, the components of $\Phi ^\lambda $ turn out to be
\begin{equation}
\Phi _n^0=\varphi _n^\alpha \varphi _{n+1}^\alpha
,\;\;\;\;\;\bbox{ \Phi
}_n=%
{\bf v}^n-{\bf a}^n,
\end{equation}
where
\begin{equation}
{\bf v}^n\equiv \frac T{\alpha ^2}{\bf V}^n=\varphi _n^0\bbox{
\varphi
}%
_{n+1}-\varphi _{n+1}^0\bbox{ \varphi }_n,\;\;\;{\bf a}^n\equiv
\frac T{\alpha ^2}{\bf A}^n=\bbox{ \varphi }_{n+1}\times \bbox{
\varphi }_n.
\end{equation}
>From this it follows that $\left( \Phi _n^0\right) ^2$ is an invariant of
the motion, namely
\begin{equation}
\left( \Phi _n^0\right) ^2{=}_{c} \left( \Phi _{n-1}^0\right) ^2,
\end{equation}
so we may write
\begin{equation}
\varphi _n^\alpha \varphi _{n+1}^\alpha {=}_{c}      C\varepsilon _n,\,
\end{equation}
for some real constant $C$ and where $\varepsilon _n=\pm 1.$ It turns out
that
\begin{equation}
C^2+{\bf v}_n^2+{\bf a}_n^2=1,
\end{equation}
which means that $-1\leq C\leq 1.\;$Hence the equation of motion can be
written in the form
\begin{equation}
\varphi _{n+1}^\alpha +\varphi _{n-1}^\alpha {=}_{c}  C\left(
\varepsilon _n+\varepsilon _{n-1}\right) \varphi _n^\alpha ,
\end{equation}
where the $\varepsilon _n$ are of magnitude $+1$ but otherwise arbitrary.
This arbitrariness can be traced to the use of the Lagrange multipliers $\mu
_n$ in the system function $\left( \ref{sys}\right) $ and is not a feature
that exists in the CT limit $T\rightarrow 0$.
In the special case that $\varepsilon _n=+1$ $\forall n$ then the equation
is recognized to equivalent to the {\it DT} harmonic oscillator discussed in
\cite{JAROSZKIEWICZ-97A}. Moreover, the bounds on the constant $C$ mean that
the motion is never hyperbolic. In this case it is found that
\[
\varphi _n^\alpha \varphi _n^\alpha =1,\;\;\;\;\;\varphi _{n+1}^\alpha
\varphi _n^\alpha =C,\;\;\;\;\;\varphi _{n+2}^\alpha \varphi _n^\alpha
=2C^2-1
\]
and generalizing this result gives the infinite set of invariants
\[
\varphi _{n+m}^\alpha \varphi _n^\alpha =T_m\left( C\right)
\]
where $T_m$ is a Chebyshef polynomial of Type 1 \cite{ARFKEN:68}
In terms of the $F,{\bf n}$ description, the parametrization is given by
\begin{equation}
U_n=c_n+is_n{\bf n}_n
\end{equation}
where $c_n\equiv \cos F_n,\;s_n\equiv \sin F_n,\;{\bf n}_n\cdot {\bf n}_n=1,$
and then
\begin{equation}
{\bf a}^n=s_ns_{n+1}{\bf n}_n\times {\bf n}_{n+1},\;\;\;{\bf v}^n=s_nc_{n+1}%
{\bf n}_n-s_{n+1}c_n{\bf n}_{n+1}.
\end{equation}
\section{The $\sigma $ model}
\subsection{continuous time}
The model is now extended to include spatial dependence, but not the quartic
terms in the original Lagrangian. We shall call this the $\sigma $ model.
The {\it CT} Lagrange density is now
\begin{equation}
{\cal L}=\frac{_1}{^4}\alpha ^2Tr\,\partial _\mu U^{+}\partial ^\mu U,
\end{equation}
which is equivalent to
\begin{equation}
{\cal L}=\frac{{}_1}{{}^2}\alpha ^2\partial _\mu \varphi ^\alpha \partial
^\mu \varphi ^\alpha +\frac{{}_1}{{}^2}\mu \left( \varphi ^\alpha \varphi
^\alpha -1\right) .
\end{equation}
The equations of motion are
\begin{equation}
\alpha ^2\Box \varphi ^\alpha =_c\mu \varphi ^\alpha ,\;\;\;\;\;\;\;\varphi
^\alpha \varphi ^\alpha =_c1
\end{equation}
which reduce to
\begin{equation}
\Box \varphi ^\alpha =_c\left( \varphi ^\beta \Box \varphi ^\beta \right)
\varphi ^\alpha .
\end{equation}
The conserved energy-momentum tensor density is
\begin{equation}
{\cal T}^{\mu \nu }=\alpha ^2\left\{ \partial ^\mu \varphi ^\alpha \partial
^\nu \varphi ^\alpha -\frac{{}_1}{{}^2}\eta ^{\mu \nu }\partial _\beta
\varphi ^\alpha \partial ^\beta \varphi ^\alpha \right\} .
\end{equation}
The invariance of the Lagrange density to the same transformation as before
allows the vector and axial currents to be determined. Under the
infinitesimal transformation
\begin{equation}
U\rightarrow U^{\prime }\equiv (1+i{ \bbox{ \tau} \cdot }{\bf
a)}U\left( 1-i%
{ \bbox{ \tau} \cdot }{\bf b}\right)
\end{equation}
then the fields change according to the rule
\begin{equation}
\delta \varphi ^0=\left( {\bf b-a}\right) \cdot \bbox{ \varphi
,\;\;\;}\delta \bbox{ \varphi }=\left( {\bf a-b}\right) \varphi
^0+\bbox{ \varphi }\times \left( {\bf a+b}\right) ,
\end{equation}
giving the conserved currents
\begin{equation}
{\bf V}^\mu =\varphi ^0\partial ^\mu \bbox{ \varphi }-\partial
^\mu
\varphi ^0%
\bbox{ \varphi ,\;\;\;A}^\mu =\partial ^\mu \bbox{ \varphi }\times
\bbox{
\varphi .%
}
\end{equation}
\subsection{discrete time}
The system function density for the $\sigma $-model is taken to be
\begin{eqnarray}
{\cal S}^n &=&\frac{_1}{^2}T\alpha ^2\{D_n^{+}\varphi _n^\alpha
D_n^{+}\varphi _n^\alpha -\frac{_1}{^2}\nabla \varphi _n^\alpha \cdot \nabla
\varphi _n^\alpha -\nabla \varphi _{n+1}^\alpha \cdot \nabla \varphi
_{n+1}^\alpha \}  \nonumber \\
&&+\frac{_1}{^4}T\mu _n\left( \varphi _n^\alpha \varphi _n^\alpha -1\right) +%
\frac{_1}{^4}T\mu _{n+1}\left( \varphi _{n+1}^\alpha \varphi _{n+1}^\alpha
-1\right) ,
\end{eqnarray}
which gives equation of motion
\begin{equation}
\frac{\alpha ^2}T\{\varphi _{n+1}^\alpha -2\varphi _n^\alpha +\varphi
_{n-1}^\alpha \}-T\alpha ^2\nabla ^2\varphi _n^\alpha =_cT\mu _n\varphi
_n^\alpha ,\;\;\;\;\;\varphi _n^\alpha \varphi _n^\alpha =_c1.
\end{equation}
There is invariance of the system function density under the infinitesimal
transformation $\left( \ref{456}\right) ,$ and this gives two conserved
currents:
\begin{equation}
D_n^{+}{\bf V}_n^0+\partial _i{\bf V}_n^i=_c{\bf 0,}\;\;\;D_n^{+}{\bf A}%
_n^0+\partial _i{\bf A}_n^i=_c{\bf 0,}
\end{equation}
where
\begin{eqnarray}
{\bf V}_n^0 &\equiv &\varphi _n^0\bbox{ \Pi }_n-\bbox{ \varphi
}_n\Pi _n^0,
\nonumber \\
{\bf V}_n^i &\equiv &\frac 1T\left\{ -\bbox{ \varphi
}_n\frac{\partial
{\cal S}%
^n}{\partial \partial _i\varphi _n^0}-\bbox{ \varphi
}_{n+1}\frac{\partial {\cal S}^n}{\partial \partial _i\varphi
_{n+1}^0}+\varphi _n^0\frac{\partial {\cal S}^n}{\partial \partial
_i\bbox{ \varphi }_n}+\varphi
_{n+1}^0\frac{%
\partial {\cal S}^n}{\partial \partial _i\bbox{ \varphi }_{n+1}}\right\}
\nonumber \\
{\bf A}_n^0 &\equiv &\bbox{ \Pi }_n\times \bbox{ \varphi }_n
\nonumber \\
{\bf A}_n^i &\equiv &\frac 1T\left\{ \frac{\partial {\cal S}^n}{\partial
\partial _i\bbox{ \varphi }_n}\times \bbox{ \varphi }_n+\frac{\partial
{\cal S}^n%
}{\partial \partial _i\bbox{ \varphi }_{n+1}}\times \bbox{ \varphi }%
_{n+1}\right\}
\end{eqnarray}
i.e
\begin{eqnarray}
{\bf V}_n^0 &\equiv &\frac{\alpha ^2}T\left\{ \varphi _n^0\bbox{
\varphi
}%
_{n+1}-\bbox{ \varphi }_n\varphi _{n+1}^0\right\}
+\frac{_1}{^2}T\alpha ^2\left\{ \bbox{ \varphi }_n\nabla ^2\varphi
_n^0-\varphi _n^0\nabla
^2\bbox{ %
\varphi }_n\right\}  \nonumber \\
{\bf V}_n^i &\equiv &\frac{_1}{^2}\alpha ^2\left\{ \bbox{ \varphi
}_n\partial _i\varphi _n^0-\varphi _n^0\partial _i\bbox{ \varphi
}_n+\bbox{
\varphi }%
_{n+1}\partial _i\varphi _{n+1}^0-\varphi _{n+1}^0\partial
_i\bbox{
\varphi }%
_{n+1}\right\}  \nonumber \\
{\bf A}_n^0 &\equiv &\frac{\alpha ^2}T\bbox{ \varphi }_{n+1}\times
\bbox{ %
\varphi }_n-\frac{_1}{^2}T\alpha ^2\nabla ^2\bbox{ \varphi
}_n\times
\bbox{ %
\varphi }_n  \nonumber \\
{\bf A}_n^i &\equiv &-\frac{_1}{^2}\alpha ^2\partial _i\bbox{ \varphi }%
_n\times \bbox{ \varphi }_n-\frac{_1}{^2}\alpha ^2\partial
_i\bbox{
\varphi }%
_{n+1}\times \bbox{ \varphi }_{n+1}.
\end{eqnarray}
\section{Discretization of the full model}
We may rewrite the full Skyrme Lagrange density equation $\left( 22\right)
\; $in the form
\begin{equation}
{\cal L}_s=\frac{_1}{^2}\dot{\varphi}^\alpha M^{\alpha \beta }\left(
\partial _i\varphi ^\mu \right) \dot{\varphi}^\beta -W\left( \partial
_i\varphi ^\mu \right) +\frac{_1}{^2}\mu \left( \varphi ^\alpha \varphi
^\alpha -1\right) ,  \label{890}
\end{equation}
where $M^{\alpha \beta }\left( \partial _i\varphi ^\mu \right) $ is given by
equation $\left( 24\right) $ and the potential function $W\left( \partial
_i\varphi ^\mu \right) $ contains quadratic and quartic terms in the
derivatives of the fields. A serious problem may arise if any term in a
Lagrangian is greater than quadratic in the dynamical variables. It is
possible in such a case that certain discretizations leads to implicit
equations of motion which cannot be solved directly to give the future
values of the variables from a knowledge of their previous values.
This remark applies to a classical theory. It is interesting that
quantization may improve the situation. Suppose we had a system with
dynamical variable $x_n$, and the {\em DT} equation of motion was an
implicit one of the form
\begin{equation}
\Phi \left( x_{n-1},x_n,x_{n+1}\right) =_c0,\;\;\;\;n=1,2,\ldots
\end{equation}
where $\Phi $ is some function. If we could not solve this equation, viz,
find a unique expression of the form
\begin{equation}
x_{n+1}=_c\Omega \left( x_{n-1},x_n\right) ,
\end{equation}
then the implication is that there exists more than one solution consistent
with the initial conditions. Alternatively, given $x_{n-1}$ and $x_{n+1}$,
there would not be a unique {\em DT }trajectory (i.e., unique value of $x_n)$
connecting these points. This is of course unsettling and contrary to
classical principles, but not so in quantum theory. We just have to recall
Feynman's path integral, which explicitly requires us to consider all
possible trajectories between initial and final times, including
non-classical ones.
It will be seen from this that {\em DT} mechanics should accommodate quantum
mechanics better than classical mechanics, except in simple cases such as
the {\em DT} harmonic oscillator. The {\em DT} Feynman rules for a $\varphi
^3$ scalar field theory was discussed in \cite{JAROSZKIEWICZ-97C}, and it
was found that the momentum space rules had softened vertices, due to the
effects of temporal discretization, which is a potentially useful result.
Analogous effects are expected from non-commutative spacetime theories.
There are several possible discretizations of the Lagrange density $\left(
\ref{890}\right) ,$ such as
\begin{eqnarray}
{\cal S}^n &=&\frac{_1}{^2}TD_n^{+}\varphi _n^\alpha M^{\alpha \beta }\left(
\frac{_1}{^2}\partial _i\varphi _n^\mu +\frac{_1}{^2}\partial _i\varphi
_{n+1}^\mu \right) D_n^{+}\varphi _n^\beta -TW\left( \frac{_1}{^2}\partial
_i\varphi _n^\mu +\frac{_1}{^2}\partial _i\varphi _{n+1}^\mu \right)
\nonumber \\
&&+\frac{_1}{^4}T\mu \left( \varphi _n^\alpha \varphi _n^\alpha -1\right) +%
\frac{_1}{^4}T\mu \left( \varphi _{n+1}^\alpha \varphi _{n+1}^\alpha
-1\right)
\end{eqnarray}
or
\begin{eqnarray}
{\cal S}^n &=&\frac{_1}{^4}TD_n^{+}\varphi _n^\alpha \left\{ M^{\alpha \beta
}\left( \partial _i\varphi _n^\mu \right) +M^{\alpha \beta }\left( \partial
_i\varphi _{n+1}^\mu \right) \right\} D_n^{+}\varphi _n^\beta   \nonumber \\
&&-\frac{_1}{^2}T\left\{ W\left( \partial _i\varphi _n^\mu \right) +W\left(
\partial _i\varphi _n^\mu \right) \right\}   \nonumber \\
&&+\frac{_1}{^4}T\mu \left( \varphi _n^\alpha \varphi _n^\alpha -1\right) +%
\frac{_1}{^4}T\mu \left( \varphi _{n+1}^\alpha \varphi _{n+1}^\alpha
-1\right) .
\end{eqnarray}
Whichever form is chosen, the (implicit) equation of motion will be given by
formula $\left( 3\right) .$
In any case, the symmetries discussed in previous sections will hold,
because the variations are global. The conserved {\em DT} vector and axial
charges can be readily worked out, and left as an exercise. There is now a
guarantee that these are dynamical invariants, even though it may not be
possible to put the equation of motion into explicit form.
\section{Discrete time energy}
The methods discussed above permit the construction of those invariants
associated with continuous symmetries, such as linear and angular momentum,
and various charges, but not energy. Fortunately, there is a way of
extending the formalism to generate a {\em DT} analogue of energy. This was
discussed by T.D. Lee \cite{LEE-83}. The method is to take the time
intervals $T_n\equiv t_{n+1}-t_n$ to be dynamical, i.e., subject to their
own dynamical equations of motion.
The first thing is to change the system function density according to the
rule
\begin{equation}
{\cal S}^n\equiv {\cal S}\left( \varphi _n^\alpha ,\varphi _{n+1}^\alpha
,\nabla \varphi _n^\alpha ,\nabla \varphi _{n+1}^\alpha ,T\right)
\rightarrow {\cal S}^n\left( T_n\right) \equiv {\cal S}\left( \varphi
_n^\alpha ,\varphi _{n+1}^\alpha ,\nabla \varphi _n^\alpha ,\nabla \varphi
_{n+1}^\alpha ,T_n\right) ,
\end{equation}
where $T_n$ is the temporal measure\ between nodes $n$ and $n+1.$ The action
integral now becomes
\begin{equation}
A_{MN}\left[ \Gamma \right] =\sum_{n=M}^{N-1}{\cal S}^n\left( T_n\right)
-\lambda \left( \sum_{n=M}^{N-1}T_n-T_{MN}\right) ,
\end{equation}
where the Lagrange multiplier enforces the constraint that the sum of the
temporal measures adds up to the fixed, total time $T_{MN}.$
The equations of motion for the ordinary dynamical variables follows exactly
the same pattern as in the regular case (constant $T),$ but now there are
the additional equations
\begin{equation}
\frac \partial {\partial T_n}{\cal S}^n\left( T_n\right) =_c\lambda
,\;\;\;n=M,M+1,...,N-1.
\end{equation}
This gives a guarantee that the object
\begin{equation}
C^n\equiv \frac \partial {\partial T_n}{\cal S}^n\left( T_n\right)
\end{equation}
is a dynamical invariant. Actually, the correct way to see what is happening
is to see the equality of the $C^n$ as dynamical equations which the $T_n$
must satisfy.
In general, numerical analysis would be required to solve these extended
equations, and this may be non-trivial. Formally, however, the problem of
energy drift is completely solved by this technique.
Finally, this approach does not cause a problem with other dynamical
invariants, such as the vector and axial charges.
\section{Quantization}
Quantization in {\em DT }particle and field theories can be readily
discussed using the Schwinger action principle, and it should be possible to
extend this method to the Skyrme Lagrangian discussed here. This will be
left for another occasion.
\section{Acknowledgements}
We thank the EPSRC for a visiting fellowship which enabled this work to be
carried out.
\section{{\protect\Large Appendix A} : {\protect\Large quaternions}}
Given the Pauli algebra
\[
\tau ^i\tau ^j=\delta _{ij}I_2+i\epsilon _{ijk}\tau ^k
\]
where $I_2$ is the $2\times 2$ identity matrix, define the quaternionic
symbols
\begin{equation}
q^i\equiv -i\tau ^i,\;\;\;q^0\equiv I_2.
\end{equation}
These satisfy the quaternionic multiplication rule
\begin{equation}
q^iq^j=-\delta _{ij}q^0+\epsilon _{ijk}q^k.  \label{34}
\end{equation}
Now define upper and lower quaternion indices with conjugation as follows:
\begin{eqnarray}
q^\mu \equiv \left( q^0,q^i\right) ,\;\; &&\;q_\mu \equiv \left(
q_0,q_i\right)  \nonumber \\
q_0\equiv q^0\;\;\;\;\;\; &&q_i\equiv -q^i=(q^i)^{*}.
\end{eqnarray}
Then conjugation is equivalent to the following quaternionic index raising
and lowering action:
\begin{equation}
\left( q^\mu \right) ^{*}=q_\mu ,\;\;\;\left( q_\mu \right) ^{*}=q^\mu .
\end{equation}
The product rule $\left( \ref{34}\right) $ may be written in the compact
form
\begin{equation}
q^\mu q^\nu =c_{\;\;\;\lambda }^{\mu \nu }q^\lambda  \label{67}
\end{equation}
where
\begin{eqnarray}
c_{\;\;\;\;0}^{00}
&=&1,\;\;\;c_{\;\;\;0}^{0i}=c_{\;\;\;0}^{i0}=0,\;\;\;c_{\;\;\;0}^{ij}=-%
\delta _{ij}  \nonumber \\
c_{\;\;\;i}^{00} &=&0,\;\;\;\;c_{\;\;\;j}^{0i}=c_{\;\;\;j}^{i0}=\delta
_{ij},\;\;\;c_{\;\;\;k}^{ij}=\epsilon _{ijk}.
\end{eqnarray}
Note that $c_{\;\;\;0}^{\mu \nu }=\eta ^{\mu \nu }=\eta _{\mu \nu },$ where $%
\eta ^{\mu \nu }$ are the components of the Lorentz metric tensor.
The product of three quaternions is given by
\begin{equation}
q^\mu q^\nu q^\alpha =c_{\;\;\;\lambda }^{\mu \nu }c_{\;\;\;\beta }^{\lambda
\alpha }q^\beta
\end{equation}
and similarly for higher powers of the quaternions.
The Pauli matrices are $2\times 2$ matrices and this permits us to define a
linear mapping $Tr,$ called the {\it trace, }on the quaternions to the reals
defined by
\begin{eqnarray}
Trq^\mu &=&2\delta _0^\mu ,\;\;Trq^\mu q^\nu =2\eta ^{\mu \nu }  \nonumber \\
Trq^0q^i &=&0,\;\;\;\;\;Trq^iq^j=-2\delta _{ij}.
\end{eqnarray}
\newpage
The trace operation and the product rule $\left( \ref{67}\right) $
permits all $SU(2)$ expressions to be simplified in terms of the
quaternions rather than $2\times 2$ matrices.

\end{document}